\documentclass[runningheads]{llncs}

\usepackage[utf8]{inputenc} 
\usepackage[T1]{fontenc}    
\usepackage[colorlinks, linkcolor=blue]{hyperref}      
\usepackage{url}            
\usepackage{booktabs}       
\usepackage{amsfonts}       
\usepackage{nicefrac}       
\usepackage{graphicx}
\usepackage[english]{babel}
\usepackage{doi}
\usepackage{amsmath}
\usepackage{listings}
\usepackage{xcolor}
\usepackage[inkscapelatex=false]{svg}
\usepackage{fontawesome5}
\usepackage{marvosym}
\usepackage{lipsum}
\newcommand \footnoteONLYtext[1]
{
    \begingroup
    \renewcommand\thefootnote{}\footnote{#1}
    \addtocounter{footnote}{-1}
    \endgroup

}
\begin{document}

\title{StenUNet: Automatic Stenosis Detection from X-ray Coronary Angiography}


\author{ 
Hui Lin \inst{1}\textsuperscript{(\Letter)}\orcidID{0000-0002-6559-2751}\and 
Tom Liu \inst{2, 3}\orcidID{0000-0003-2351-236X} \and 
Aggelos Katsaggelos\inst{1}\orcidID{0000-0003-4554-0070} \and
Adrienne Kline\inst{2,3}\orcidID{0000-0002-0052-0685}
}
\authorrunning{H. Lin et al.}
%
\institute{
Department of Electrical and Computer Engineering, Northwestern University, Evanston, IL 60208, USA \and 
Center for Artificial Intelligence, Bluhm Cardiovascular Institute, Northwestern Medicine, Chicago IL 60611, USA \and
Department of Cardiac Surgery, Northwestern University, Chicago IL 60611, USA\\
\email{\textsuperscript{\Letter}huilin2023@u.northwestern.edu}\\
}

\maketitle              
\footnoteONLYtext{H. Lin and T. Liu—The authors contributed equally to this work.}
\begin{abstract}

Coronary angiography continues to serve as the primary method for diagnosing coronary artery disease (CAD), which is the leading global cause of mortality. The severity of CAD is quantified by the location, degree of narrowing (stenosis), and number of arteries involved. In current practice, this quantification is performed manually using visual inspection and thus suffers from poor inter- and intra-rater reliability. The MICCAI grand challenge: Automatic Region-based Coronary Artery Disease diagnostics using the X-ray angiography imagEs (ARCADE) curated a dataset with stenosis annotations, with the goal of creating an automated stenosis detection algorithm. Using a combination of machine learning and other computer vision techniques, we propose the architecture and algorithm StenUNet to accurately detect stenosis from X-ray Coronary Angiography. Our submission to the ARCADE challenge placed 3rd among all teams. We achieved an F1 score of 0.5348 on the test set, 0.0005 lower than the 2nd place.

\end{abstract}
%
%
\section{Introduction}
\label{Introduction}
A coronary angiogram is a medical procedure used to diagnose and evaluate coronary artery disease (CAD), where the blood vessels that supply oxygen and nutrients to the heart muscle become narrowed or blocked. The procedure involves injecting a contrast dye into the coronary arteries and taking X-ray images to visualize the blood flow within coronary arteries. Stenosis detection during a coronary angiogram is one critical aspect of the procedure, as it helps assess the degree of narrowing in these arteries and thus disease severity and the planning of subsequent interventions  \cite{de1991quantitative}.

Stenosis refers to the narrowing of the coronary arteries and is most commonly due to the buildup of plaque (atherosclerosis). The degree of stenosis is measured as a percentage of the artery's diameter that is blocked (e.g. 70\% is considered significant blockage \cite{fleming1991patterns}). Stenosis detection is primarily based on a visual assessment by experienced interventional cardiologists, who examine these X-ray images to identify areas of stenosis and quantify their severity and anatomical locations. The limitation of solely visual inspection for disease severity determination is that it requires significant expertise and manual effort. Moreover, as a result of the manual approach, there exists a high degree of intra- and inter-observer variability \cite{galbraith1978coronary,leape2000effect,zir1976interobserver}. Therefore, there is a strong interest in achieving automatic stenosis detection with reduced time, improved reproducibility, and the potential for higher accuracy. 

Automatic stenosis detection from X-ray coronary angiography is a challenging task due to the varied and complex anatomy of arteries, suboptimal image quality, and the relatively small size of stenotic regions. Considerable efforts have been devoted to overcoming the aforementioned challenges. Given that stenotic areas are characterized by significant changes in radius, some methods have devised multi-stage frameworks. These frameworks initially focus on vessel extraction and diameter estimation before identifying the stenosis area. For instance, a framework proposed by Wan et al. \cite{wan2018automated} is composed of Hessian-based vessel enhancement, level-set skeletonization, vessel diameter estimation, and local extremum identification. These multi-stage frameworks are susceptible to error propagation, which can make optimization and parameter tuning challenging. Fortunately, convolutional neural networks (CNNs) have demonstrated promising performance in optimization and parameter tuning because of their inherent ability to automatically learn and adapt to complex patterns and features within data \cite{lin2019automated,mao2023deep,mozaffar2021geometry}. 

CNN-based approaches for automatic stenosis detection from X-ray coronary angiography can be categorized into single-frame and multi-frame methods. The single-frame methods analyze each frame independently \cite{ovalle2022hybrid,cong2019automated,wan2018automated}. A hybrid network named H-CQN \cite{ovalle2022hybrid} combines a quantum computing technique with a classic CNN. The features extracted by the classical network are mapped into a hypersphere by a hyperbolic tangent function and then fed into the following quantum network. It performed binary classification (stenosis or non-stenosis) on patches with a fixed size obtained from XCA images. These historical methods do not take into account the temporal information within the X-ray coronary angiography (XCA) sequence, and as a result, they may encounter challenges when handling difficult frames that exhibit poor image quality.

Some previous work leverages sequential and/or aggregated temporal information from XCA for analysis \cite{pang2021stenosis,han2023coronary}. Employing features from neighboring frames can provide valuable support for handling difficult frames, improving the robustness and accuracy of stenosis detection in adverse conditions. A CNN-based network named Stenosis-DetNet \cite{pang2021stenosis} leveraged sequential information via feature fusion and sequence consistency. They demonstrated, that utilizing temporal information across sequential frames, enhanced the accuracy of stenosis detection. This was achieved via the fusion of all frames' features for the frame-level classification and regression of its candidate boxes. Detected boxes from each frame are considered prior to filtering. Boxes with small frequencies are removed, and any missed boxes are interpolated using coronary structure constraints. Alternatively, a transformer-based framework is proposed to model the long-range spatio-temporal context \cite{han2023coronary}. Here, visual tokens are obtained by the proposal-shifted spatio-temporal tokenization (PSSTT), which are further fed into a transformer-based feature aggregation (TFA) network to enhance the extracted features for final stenosis detection. 

To date, these prior methods have been exclusively developed and validated on their respective unpublished datasets, and there hasn't been any comparative research to assess their performance against one another on a public dataset. Further, it's crucial to underline that the stenosis localization achieved by these methods is not as precise as pixel-wise segmentation \cite{ovalle2022hybrid,cong2019automated,wan2018automated,pang2021stenosis,han2023coronary}. Some methods employ bounding boxes, while others generate attention areas using class activation maps (CAM). Thus, although the general positions of stenosis may be detected, the exact positions and extent of the stenosis remain unclear. 
 
To overcome the shortcomings of the methods mentioned above, we propose a 2D UNet-based model, StenUNet\footnote[1]{The code is available at \href{https://github.com/HuiLin0220/StenUNet}{https://github.com/HuiLin0220/StenUNet}}, to achieve pixel-wise stenosis detection from single frame X-ray coronary angiography. Of note, multi-frame temporal information was not available. The contributions of this paper are summarized as follows:

\begin{enumerate}
  \item A UNet-based network, StenUNet, is introduced for pixel-wise stenosis localization, offering enhanced accuracy in determining the exact position and size of stenotic regions. This advancement holds significant potential for assisting in the treatment planning of coronary artery disease in clinical applications, where precise localization and extent are crucial for effective medical intervention.
  \item This work culminates in a small network and computationally efficient pipeline, making it deployable in real-world edge applications.
\end{enumerate}

\section{Methods}
\label{Methods}

The pipeline and the proposed StenUNet architecture are described in this section. The loss function applied for training StenUNet is the combination of binary cross-entropy loss and dice loss, which are also outlined in this section.

\subsection{Preprocess and Postprocessing}
\label{Preprocess}

\begin{figure*}[ht]
    \centering
    
    \includegraphics[width=0.8\textwidth]{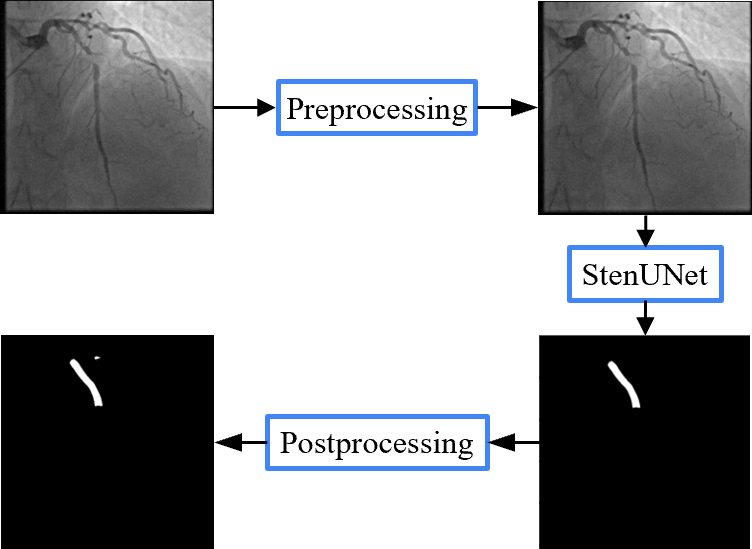}
    \caption{The proposed pipeline for stenosis detection from X-ray angiography. An example is from the testing set. StenUNet's architecture is depicted in Fig.\ref{fig: StenUNet}.}
    \label{fig: pipeline}
\end{figure*}

As shown in Fig.\ref{fig: pipeline}, each angiogram undergoes preprocessing before being fed into StenUNet. Subsequently, the output from StenUNet undergoes post-processing to generate the final prediction.

\noindent \\ 
Preprocessing steps include homomorphic image enhancement performed with a Butterworth high pass filter to reduce background noise and glare as described in \cite{ddfb_khan}. Briefly, in Eq. (\ref{equation: illuminance}) an input image $I$ can be expressed as the sum of its log-transformed illuminance $ln(i)$ and reflectance $ln(r)$ in the Fourier domain. A Butterworth low pass filter of order $n$, and with cutoff frequency $D_0$ and $D(u,v)=\sqrt(u^2+v^2)$, with $(u,v)$ the discrete frequency domain coordinates, is applied (\ref{equation: butterworth}, \ref{equation: homomorphic}) in order to reduce noise.

\begin{equation}
\label{equation: illuminance}
\ln\{I(x, y)\} = \ln\{i(x, y)\} + \ln\{r(x, y)\}
\end{equation}

\begin{equation}
\label{equation: butterworth}
H(u, v) = \frac{1}{1 + \left(\frac{D(u, v)}{D_0}\right)^{2n}}
\end{equation}

\begin{equation}
\label{equation: homomorphic}
N(u,v)=H(u,v) \times {\cal{F}}\{\ln\{ I(x, y) \} \}
\end{equation}
where $\cal{F}$ denotes the 2D discrete Fourier transform.

The resultant image is normalized using Eq. \ref{equation: normalization} according to \cite{ddfb_khan}.

\noindent \\ 
\begin{equation}
\label{equation: normalization}
N(i, j) =
\begin{cases}
    M_0 + \sqrt{\frac{{\text{VAR}_0 \cdot (I_{En} - M)^2}}{{\text{VAR}}}} & \text{if } I_{En}(i, j) > M \\
    M_0 - \sqrt{\frac{{\text{VAR}_0 \cdot (I_{En} - M)^2}}{{\text{VAR}}}} & \text{otherwise}
\end{cases}
\end{equation}
where $M$ and VAR indicate the mean and variance of
input image. $M_{0}$ and $\text{VAR}_{0}$ are desired mean and variance. $N$ is the normalized image.

StenUNet's predictions sometimes include small segments that do not correspond to coronary anatomy and, therefore, are not stenosis. The final refinement of these predictions is achieved through a post-processing method specifically designed to eliminate these small segments. The enhancements resulting from the preprocessing and post-processing steps are detailed in Section \ref{Ablation}.

\subsection{The network architecture}
\label{The architecture}

UNet-based methods are famous for their computationally efficient architectures that simultaneously offer high performance in medical image analysis. The skip connections inside UNet not only recover spatial information for fine-grained segmentation but also alleviate the potential vanishing gradient problem during training. Therefore, StenUNet, a UNet-based network, is proposed and implemented on the nnU-Net framework \cite{isensee2021nnu}. StenUNet takes preprocessed X-ray angiography as its input and applies additional pre- and post-processing steps.

\begin{figure*}[t]
    \centering
    \includegraphics[width=\textwidth]{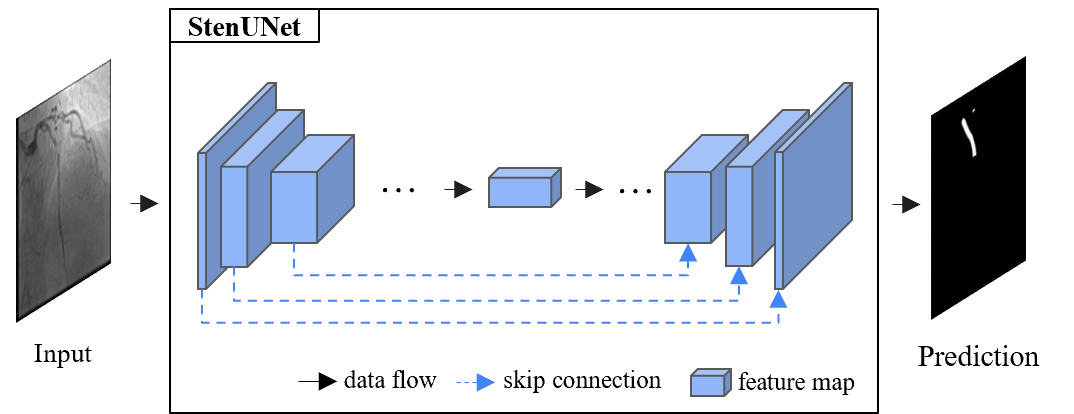}
    \caption{The proposed UNet-based StenUNet for stenosis detection. The reader is referred to Section \ref{Methods} for more details on the StenUNet algorithm.}
    \label{fig: StenUNet}
\end{figure*}

As depicted in Fig. \ref{fig: StenUNet}, StenUNet is composed of an encoder and decoder network on the left and right sides, respectively. The encoder network extracts high-level features from the input while progressively reducing the size of feature maps to expand the receptive field. Subsequently, the decoder network gradually reconstructs these features to generate segmentation maps at the original size. Skip connections link the high-level and low-level features.

\subsection{Loss Function}
\label{Loss Function}

While training StenUNet, the difference between the output and the ground truth is calculated using Eq. \ref{equation: seg loss}. The total segmentation loss is defined as $\mathcal{L}$, which is the weighted sum of binary cross entropy loss (BCE) ($\mathcal{L}^{BCE}$) and dice loss ($\mathcal{L}^{dice}$) \cite{galdran2022optimal}. The BCE loss is applied to achieve a stable and effective training process, while the dice loss is applied to address and 
eliminate the effect of class imbalance between the stenosis and the background. $\mathcal{L}$ is given by:

\begin{equation}
\label{equation: seg loss}
\begin{split}
    \mathcal{L} &= \mathcal{L}^{BCE}  + \mathcal{L}^{dice}\\
    &= -(Y\text{log}\hat{Y}+(1-Y)\text{log}(1-\hat{Y}))+ (1-\frac{2Y\cap \hat{Y}}{Y\cup \hat{Y}})
\end{split}
\end{equation}
where $Y$ is the ground truth,  $Y\in \{0, 1\}$ and $\hat{Y}$ is the model output, $\hat{Y}\in [0, 1]$, where 0 represents the background, and 1 represents pixel-level stenosis.

\section{Experiments}
\label{Experiments}

The proposed method is validated in the Automatic Region-based Coronary Artery Disease diagnostics using x-ray angiography imagEs (ARCADE) challenge dataset\footnote[2]{\href{https://arcade.grand-challenge.org/} {https://arcade.grand-challenge.org/}} \cite{maxim_popov_2023_7981245}. In this section, we delve into the stenosis detection task.

\subsection{ARCADE Challenge Dataset}
\label{Dataset}

\begin{figure}[ht]
    \centering
    \includegraphics[width=\columnwidth]{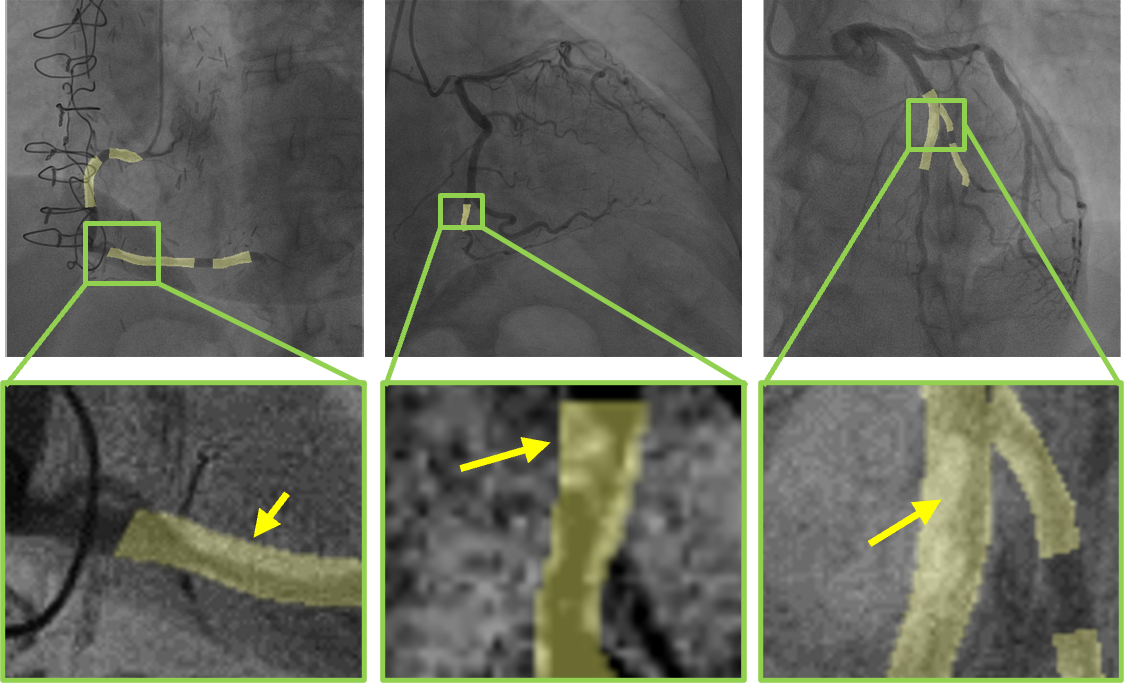}
    \caption{Examples of 2D X-ray coronary angiography and their corresponding annotated stenosis regions. The yellow areas represent annotated stenosis overlaid on the X-ray images. In the second row, green bounding boxes offer magnification of these structures for enhanced visualization. Yellow arrows denote the annotations that include not only the narrowed artery but also the inclusion of surrounding tissues.
}
    \label{fig: example}
\end{figure}

The ARCADE challenge dataset \cite{maxim_popov_2023_7981245} is provided by the Research Institute of Cardiology and Internal Diseases and CMC Technologies LLP. It includes 1000 static 2-dimensional X-ray images for training, 200 for validation, and  300 for testing. All images are of size 512$\times$512 pixels. Regions containing atherosclerotic plaques were carefully annotated by medical experts, and are represented as polygons as shown in Fig. \ref{fig: example}.

Polygon annotations represent an intermediate level of detail compared to pixel-wise annotations and bounding boxes. Polygons offer a more precise delineation of the stenosis region than bounding boxes, as they can more closely approximate the true shape of the stenotic artery segment. However, they are less granular than pixel-wise annotations, which involve marking each individual pixel within the stenosis region. 

When analyzing Fig \ref{fig: example}, automatic stenosis detection from X-ray coronary angiography faces the following challenges: 

(1) Calcified plaque: Coronary artery stenosis can involve calcified plaque deposits. These calcifications may appear similar to blood vessels on X-ray images, potentially leading to false positives or making it challenging to distinguish between the two. 

(3) Image quality and noise: X-ray angiography images can be noisy, and the quality may vary depending on factors like the patient's anatomy, the contrast agent's distribution, and the X-ray equipment used. These can all decrease the signal-to-noise ratio (SNR). This noise can make it difficult for computer vision algorithms to accurately identify and quantify stenosis.

(4) Class imbalance: As the stenosis is a tiny part of the whole image. When training StenUNet on imbalanced datasets, algorithms tend to become biased towards the majority class (non-stenosis). As a result, they may have difficulty correctly identifying the minority class (stenosis).

(5) Polygon annotations: These denote stenotic and non-stenotic areas, which are not as precise as pixel-wise annotations, as highlighted by yellow arrows in Fig \ref{fig: example}. This can confound the model during training.

\subsection{Implementation details and Metrics}
\label{Metrics}

All experiments were conducted on a workstation equipped with a single NVIDIA A100-PCI GPU card with 40 GB memory. The StenUNet model was trained using the Stochastic Gradient Descent (SGD) optimizer. For the learning rate schedule, a polynomial decay learning rate schedule was adopted, with an initial learning rate set to 0.01. The use of a polynomial decay learning rate schedule offered the advantage of achieving stable convergence by gradually reducing the learning rate over time, thereby enabling better fine-tuning and improved final performance.

Data augmentation was employed to prevent overfitting and overcome class imbalance mentioned in Section \ref{Dataset}. Two spatial transformation methods; scaling and rotation, were applied to each XY plane with a probability of 50\% for data augmentation. The scaling factor and rotation angle were randomly selected within $(0.7, 1.4)$, $(-180^{\circ}, 180^{\circ})$ around the X axis. Two noise transformation methods, Gaussian noise and blur, were applied. The variance of Gaussian noise was randomly chosen from the interval (0, 0.1). The kernel size for Gaussian blur remained fixed at 3, while the variance of Gaussian blur was randomly selected from the range (0.5, 1).

The encoder of StenUNet comprises seven convolutional stages. Each convolution performed in the StenUNet uses 2D kernels with size, $3\times3$ followed by a max pooling layer using kernels with size $2\times2$. The channel numbers at all stages are, respectively, 32, 64, 128, 256, 512, 512, and 512 to gradually improve the model's representation ability.

F1 score and inference time were the two main evaluation metrics. The inference time as outlined by the challenge needed to be less than 5s/image. The average of the F1 Scores of all X-ray images are calculated for algorithm evaluation. Each individual X-ray image's F1 scores are calculated using Equation \ref{equation: F1}.

\begin{equation}
\label{equation: F1}
    \begin{split}
F1 &= \frac{2}{\frac{1}{Recal} + \frac{1}{Precision}}\\
&= \frac{2}{\frac{TP+FN}{TP} + \frac{TP+FP}{TP}}
    \end{split}
\end{equation}
where $TP$, $FN$, $TN$, and $FP$ represent the number of true positives, false negatives, true negatives, and false positives,  respectively.

\section{Results and Discussion}
\label{Results}

\subsection{Ablation study}
\label{Ablation}

Table \ref{table: ablation study of the proposed method StenUNet on the ARCADE dataset} presents the evaluation results achieved by the proposed StenUNet method with and without preprocessing or post-processing techniques applied. StenUNet-raw is the proposed pipeline without preprocessing and post-processing processes. StenUNet-pre is without post-processing. Both pre- and post-processing techniques improve StenUNet's performance in terms of F1 score by 1.26\% and 3.4\%, respectively. Together, the F1 score was improved by 4.66\%.

\begin{figure}[htbp]
    \centering
    \includegraphics[width=\columnwidth]{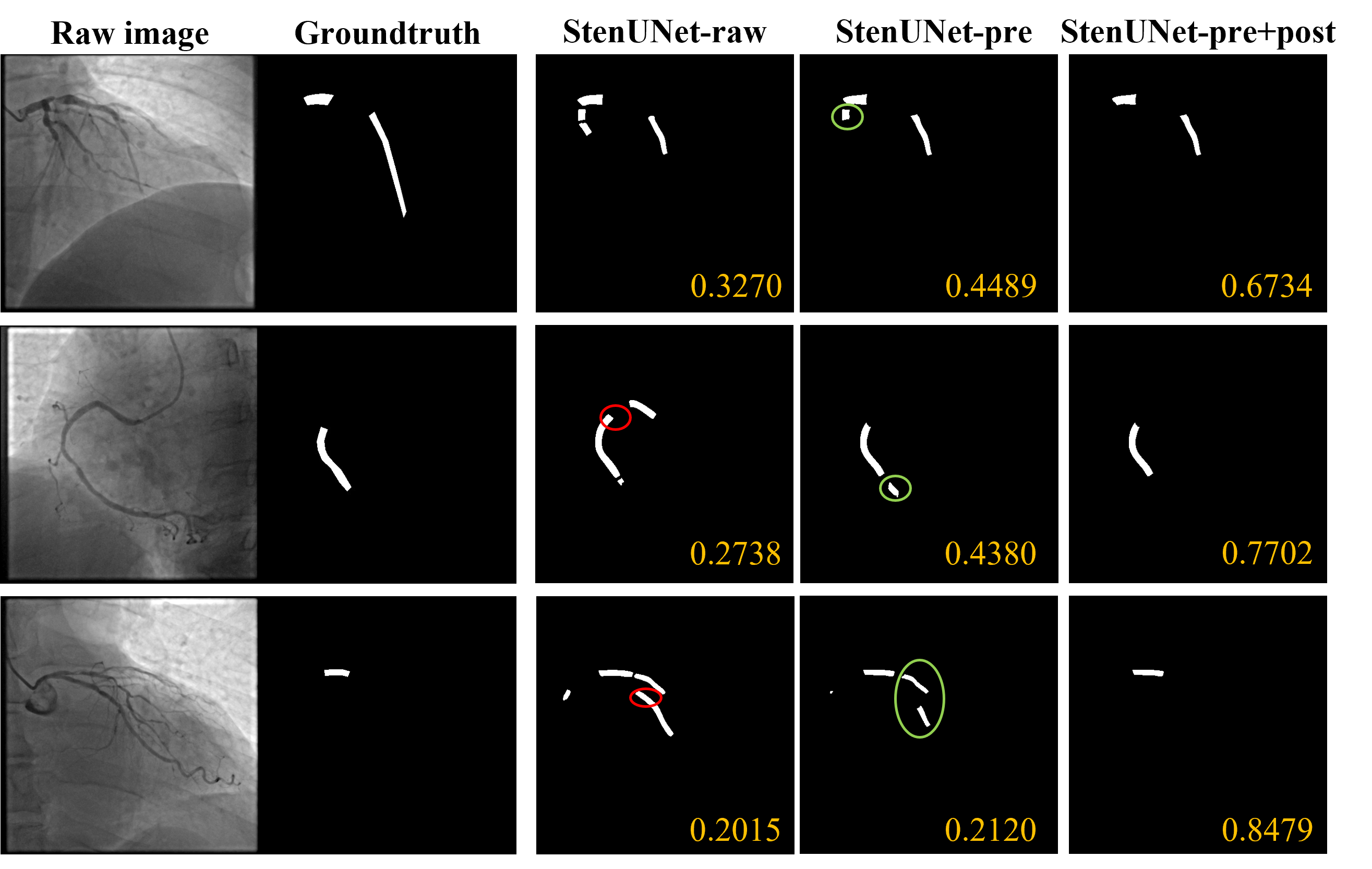}
    \caption{2D visualization of stenosis detection results by StenUNet-raw, StenUNet-pre, and StenUNet-pre+post. The orange number displayed in the bottom-right corner of each picture represents the individual image's F1 score. The leftmost two columns are the raw XCA images and their corresponding ground truth annotations.}
    \label{fig:2dvisulization_models}
\end{figure}

Fig. \ref{fig:2dvisulization_models} displays three examples of the stenosis detection results obtained by StenUNet-raw, StenUNet-pre, and StenUNet-pre+post. It is evident that preprocessing plays a crucial role in effectively eliminating false positives, as indicated by the red circles. These false positives cannot be adequately addressed solely by removing small segments. Thus, the preprocessing step is indispensable for improving the accuracy of the detection process. In comparison to the preprocessing step, post-processing assumes a more critical role by efficiently eliminating numerous, and sometimes relatively larger, small segments, leading to a substantial increase in the F1 score. False positives removed by the post-processing are indicated by the green circles.

\begin{table}[htbp]
\begin{center}
\caption{Evaluation of the ablation study performed on StenUNet on the ARCADE dataset. In StenUNet-raw, the preprocessing and post-processing processes are removed from the proposed pipeline. StenUNet-pre tests the removal of only the post-processing algorithm. The best result is highlighted in bold, and the second-best result is underlined.}
\label{table: ablation study of the proposed method StenUNet on the ARCADE dataset}
\begin{tabular}{ c ||c }
\hline
Methods & F1 (\%)\\
\hline
\hline
StenUNet-raw &  $48.34$ \\
\hline
StenUNet-pre& \underline{$49.60$}\\
\hline
\textbf{StenUNet-pre+post}
&\boldmath$53.00$ ($53.48$ on the leaderboard)\\ 
 \hline
\end{tabular}
\end{center}
\end{table}

\subsection{Error analysis}
\label{error}

\begin{figure}[htbp]
    \centering
    \includegraphics[width=\columnwidth]{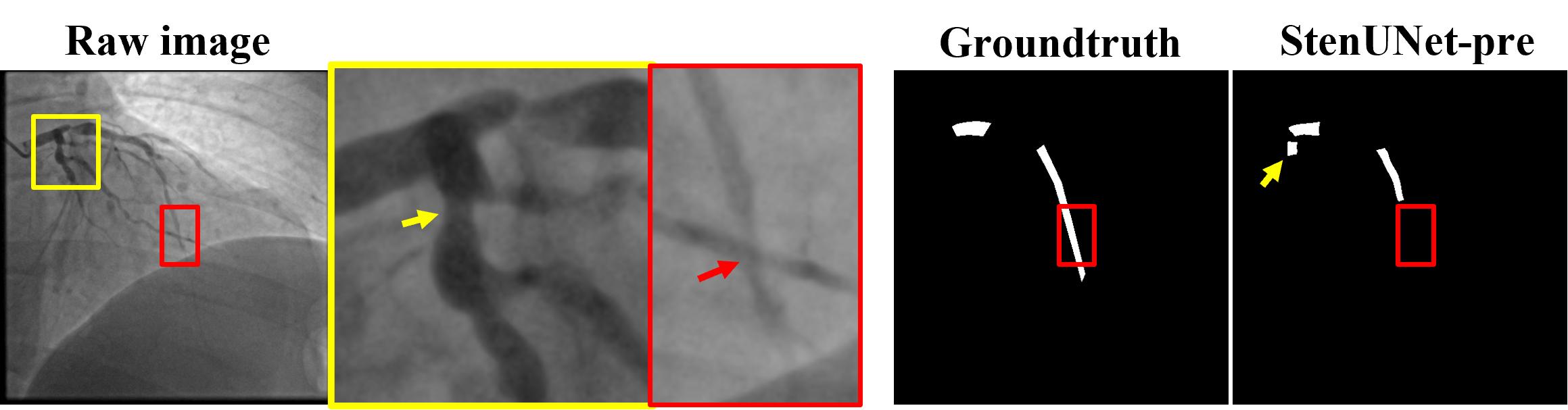}
    \caption{A zoomed-in view of the example in the first row of Fig.\ref{Results} for error analysis. The yellow and red bounding boxes indicate two areas where significant errors are present. Arrows point out the most confusing areas for models.}
    \label{fig:error}
\end{figure}

A limitation of StenUNet is that it exhibits a tendency to overlook thin and indistinct arteries. As illustrated by the red bounding box and arrow in Fig. \ref{fig:error}, the proposed StenUNet fails to detect the long segment. This is primarily due to the fact that the long segment is significantly lighter in appearance compared to other parts of the artery. It becomes challenging to distinguish it from other anatomy, let alone detect stenosis within it. As indicated by the yellow arrow, the point of intersection (vessel crossing) appears denser, while the surrounding area around the denser intersection tends to be detected incorrectly.

\section{Conclusions}
\label{Conclusions}
StenUNet offers a novel approach to automate the detection of stenosis from X-ray coronary angiograms. This method achieves pixel-wise segmentation, which is more precise compared to state-of-the-art methods whose predictions are bounding boxes. Additionally, StenUNet's computationally efficient architecture makes it highly accessible for deployment in practical real-world applications.

While StenUNet currently operates on individual frames, it's important to note that X-ray coronary angiograms are captured dynamically over a sequence of frames. Consequently, one future direction involves leveraging networks capable of handling frame sequences, such as RNNs or transformers. These networks can effectively integrate crucial information from other frames within the time series, potentially enhancing stenosis detection accuracy. Although not part of the challenge, another future direction for this work entails quantifying the degree of stenosis present in the anatomy where stenosis was detected,  which holds significant importance in clinical applications. 


\end{document}